\documentclass[12pt,titlepage,letterpaper]{utarticle}

\usepackage{amsmath}
\usepackage{mathrsfs}
\usepackage{amsfonts}
\usepackage{amssymb}
\usepackage{amsthm}
\usepackage{mathtools}
\usepackage{graphicx}
\usepackage{color}
\usepackage{ucs}
\usepackage[utf8x]{inputenc}
\usepackage{xparse}
\usepackage{tikz}
\usepackage{hyperref}

\definecolor{aqua}{rgb}{0, 1.0, 1.0}
\definecolor{fuschia}{rgb}{1.0, 0, 1.0}
\definecolor{gray}{rgb}{0.502, 0.502, 0.502}
\definecolor{lime}{rgb}{0, 1.0, 0}
\definecolor{maroon}{rgb}{0.502, 0, 0}
\definecolor{navy}{rgb}{0, 0, 0.502}
\definecolor{olive}{rgb}{0.502, 0.502, 0}
\definecolor{purple}{rgb}{0.502, 0, 0.502}
\definecolor{silver}{rgb}{0.753, 0.753, 0.753}
\definecolor{teal}{rgb}{0, 0.502, 0.502}

\makeatletter
\newdimen\itex@wd%
\newdimen\itex@dp%
\newdimen\itex@thd%
\def\itexspace#1#2#3{\itex@wd=#3em%
\itex@wd=0.1\itex@wd%
\itex@dp=#2ex%
\itex@dp=0.1\itex@dp%
\itex@thd=#1ex%
\itex@thd=0.1\itex@thd%
\advance\itex@thd\the\itex@dp%
\makebox[\the\itex@wd]{\rule[-\the\itex@dp]{0cm}{\the\itex@thd}}}
\makeatother

\makeatletter
\newif\if@sup
\newtoks\@sups
\def\append@sup#1{\edef\act{\noexpand\@sups={\the\@sups #1}}\act}%
\def\reset@sup{\@supfalse\@sups={}}%
\def\mk@scripts#1#2{\if #2/ \if@sup ^{\the\@sups}\fi \else%
  \ifx #1_ \if@sup ^{\the\@sups}\reset@sup \fi {}_{#2}%
  \else \append@sup#2 \@suptrue \fi%
  \expandafter\mk@scripts\fi}
\def\tensor#1#2{\reset@sup#1\mk@scripts#2_/}
\def\multiscripts#1#2#3{\reset@sup{}\mk@scripts#1_/#2%
  \reset@sup\mk@scripts#3_/}
\makeatother

\makeatletter
\newbox\slashbox \setbox\slashbox=\hbox{$/$}
\def\itex@pslash#1{\setbox\@tempboxa=\hbox{$#1$}
  \@tempdima=0.5\wd\slashbox \advance\@tempdima 0.5\wd\@tempboxa
  \copy\slashbox \kern-\@tempdima \box\@tempboxa}
\def\slash{\protect\itex@pslash}
\makeatother

\def\clap#1{\hbox to 0pt{\hss#1\hss}}

\let\oldroot\root
\def\root#1#2{\oldroot #1 \of{#2}}
\renewcommand{\sqrt}[2][]{\oldroot #1 \of{#2}}

\DeclareSymbolFont{symbolsC}{U}{txsyc}{m}{n}
\SetSymbolFont{symbolsC}{bold}{U}{txsyc}{bx}{n}
\DeclareFontSubstitution{U}{txsyc}{m}{n}

\DeclareSymbolFont{stmry}{U}{stmry}{m}{n}
\SetSymbolFont{stmry}{bold}{U}{stmry}{b}{n}

\DeclareFontFamily{OMX}{MnSymbolE}{}
\DeclareSymbolFont{mnomx}{OMX}{MnSymbolE}{m}{n}
\SetSymbolFont{mnomx}{bold}{OMX}{MnSymbolE}{b}{n}
\DeclareFontShape{OMX}{MnSymbolE}{m}{n}{
    <-6>  MnSymbolE5
   <6-7>  MnSymbolE6
   <7-8>  MnSymbolE7
   <8-9>  MnSymbolE8
   <9-10> MnSymbolE9
  <10-12> MnSymbolE10
  <12->   MnSymbolE12}{}

\makeatletter
\def\re@DeclareMathSymbol#1#2#3#4{%
    \let#1=\undefined
    \DeclareMathSymbol{#1}{#2}{#3}{#4}}
\re@DeclareMathSymbol{\neArrow}{\mathrel}{symbolsC}{116}
\re@DeclareMathSymbol{\neArr}{\mathrel}{symbolsC}{116}
\re@DeclareMathSymbol{\seArrow}{\mathrel}{symbolsC}{117}
\re@DeclareMathSymbol{\seArr}{\mathrel}{symbolsC}{117}
\re@DeclareMathSymbol{\nwArrow}{\mathrel}{symbolsC}{118}
\re@DeclareMathSymbol{\nwArr}{\mathrel}{symbolsC}{118}
\re@DeclareMathSymbol{\swArrow}{\mathrel}{symbolsC}{119}
\re@DeclareMathSymbol{\swArr}{\mathrel}{symbolsC}{119}
\re@DeclareMathSymbol{\nequiv}{\mathrel}{symbolsC}{46}
\re@DeclareMathSymbol{\Perp}{\mathrel}{symbolsC}{121}
\re@DeclareMathSymbol{\Vbar}{\mathrel}{symbolsC}{121}
\re@DeclareMathSymbol{\sslash}{\mathrel}{stmry}{12}
\re@DeclareMathSymbol{\bigsqcap}{\mathop}{stmry}{"64}
\re@DeclareMathSymbol{\biginterleave}{\mathop}{stmry}{"6}
\re@DeclareMathSymbol{\invamp}{\mathrel}{symbolsC}{77}
\re@DeclareMathSymbol{\parr}{\mathrel}{symbolsC}{77}
\makeatother

\makeatletter
\def\Decl@Mn@Delim#1#2#3#4{%
  \if\relax\noexpand#1%
    \let#1\undefined
  \fi
  \DeclareMathDelimiter{#1}{#2}{#3}{#4}{#3}{#4}}
\def\Decl@Mn@Open#1#2#3{\Decl@Mn@Delim{#1}{\mathopen}{#2}{#3}}
\def\Decl@Mn@Close#1#2#3{\Decl@Mn@Delim{#1}{\mathclose}{#2}{#3}}
\Decl@Mn@Open{\llangle}{mnomx}{'164}
\Decl@Mn@Close{\rrangle}{mnomx}{'171}
\Decl@Mn@Open{\lmoustache}{mnomx}{'245}
\Decl@Mn@Close{\rmoustache}{mnomx}{'244}
\Decl@Mn@Open{\llbracket}{stmry}{'112}
\Decl@Mn@Close{\rrbracket}{stmry}{'113}
\makeatother

\makeatletter
\DeclareRobustCommand\widecheck[1]{{\mathpalette\@widecheck{#1}}}
\def\@widecheck#1#2{%
    \setbox\z@\hbox{\m@th$#1#2$}%
    \setbox\tw@\hbox{\m@th$#1%
       \widehat{%
          \vrule\@width\z@\@height\ht\z@
          \vrule\@height\z@\@width\wd\z@}$}%
    \dp\tw@-\ht\z@
    \@tempdima\ht\z@ \advance\@tempdima2\ht\tw@ \divide\@tempdima\thr@@
    \setbox\tw@\hbox{%
       \raise\@tempdima\hbox{\scalebox{1}[-1]{\lower\@tempdima\box
\tw@}}}%
    {\ooalign{\box\tw@ \cr \box\z@}}}
\makeatother

\makeatletter
\NewDocumentCommand\mathraisebox{moom}{%
\IfNoValueTF{#2}{\def\@temp##1##2{\raisebox{#1}{$\m@th##1##2$}}}{%
\IfNoValueTF{#3}{\def\@temp##1##2{\raisebox{#1}[#2]{$\m@th##1##2$}}%
}{\def\@temp##1##2{\raisebox{#1}[#2][#3]{$\m@th##1##2$}}}}%
\mathpalette\@temp{#4}}
\makeatletter

\makeatletter
\def\udots{\mathinner{\mkern2mu\raise\p@\hbox{.}
\mkern2mu\raise4\p@\hbox{.}\mkern1mu
\raise7\p@\vbox{\kern7\p@\hbox{.}}\mkern1mu}}
\makeatother

\newcommand{\gt}{>}

\theoremstyle{plain}

\theoremstyle{definition}

\theoremstyle{remark}

\usetikzlibrary{arrows.meta,shapes,decorations.markings}

\begin{document}

\preprint{
UTTG--15--20\\
}

\title{A Note on S.~Weinberg, ``Massless Particles in Higher Dimensions"}

\author{Jacques Distler
     \oneaddress{
     Theory Group\\
     Department of Physics,\\
     University of Texas at Austin,\\
     Austin, TX 78712, USA \\
     {~}\\
      \email{distler@golem.ph.utexas.edu}
}}

\date{October 14, 2020}

\Abstract{
In \cite{Weinberg:2020nsn}, Weinberg made a conjecture about the little-group representations of massless particles that can be created out of the vacuum by the action of a local operator in $d$ dimensions, generalizing his old result \cite{Weinberg:1964ev} in $d=4$. In this note, I prove his conjecture and extend it to arbitrary irreps of $so(1,d-1)$.
}

\maketitle

\newpage
\setcounter{page}{1}


In \cite{Weinberg:2020nsn}, Steven Weinberg posed the following question. Consider a local operator, $O(x)$, transforming in some irreducible unitary (infinite-dimensional) representation of the Poincar\'e algebra $iso(1,d-1)$. Assume that $O(x)$ has a nonzero matrix element between a massless 1-particle state and the vacuum
\begin{displaymath}
0 \neq \langle 0|O(x)|k,R'\rangle
\end{displaymath}
By translational invariance, this matrix element is nonvanishing if and only if
\begin{equation}
0 \neq \langle 0|O(0)|k,R'\rangle
\label{matrixelt}\end{equation}
We may therefore assume that $O(0)$ transforms as some (nonunitary) finite-dimensional irreducible representation, $R$, of the Lorentz algebra $so(1,d-1)$ which leaves the point $x=0$ fixed. Weinberg's question is:

\begin{itemize}%
\item Given $R$, what representation, $R'$, of $so(d-2)\subset iso(d-2)$ is compatible with a nonzero matrix element \eqref{matrixelt}?
\end{itemize}
Weinberg computed some examples, and conjectured an answer for (tensorial) representations $R$, given by Young Tableaux. Here I prove his conjecture and extend it to all finite-dimensional irreps, $R$.

The problem reduces to one in Lie theory. As an irrep of $so(1,d-1)$, $R$ is \emph{a fortiori} a representation of the little algebra, $iso(d-2)\subset so(1,d-1)$. As a representation of $iso(d-2)$, $R$ is \emph{reducible} but \emph{indecomposable}.

Writing $iso(d-2) = so(d-2) \ltimes K$, where $K$ is the $(d-2)$-dimensional abelian subalgebra of $so(1,d-1)$ which (along with $so(d-2)$) leaves fixed a particular null momentum. The irreducible subrepresentation, $R' \subset R$ is such that $K$ restricts to \emph{zero} on $R'$. An alternative characterization of $R'$ is that it is \emph{simultaneously} an irrep of $iso(d-2)$ and of $so(1,1)\times so(d-2)$, where the $so(d-2)$ is the common subalgebra of these two maximal subalgebras of $so(1,d-1)$.

With this reformulation, we can ask, ``If $R'$ is an irrep of $so(1,1)\times so(d-2)$, which irrep is it?''

To answer that, we note that the highest-weight of $R$ is contained in $R'$.

\noindent
\textbf{Proof}: Since $K$ raises\footnote{For our conventions for the $so(1,1)$ weights, see \eqref{adjdecomp}.}  the $so(1,1)$ weight, it necessarily annihilates the highest weight of $R$. Acting with $so(d-2)$ does not change the $so(1,1)$ weight and the commutator of an element of $so(d-2)$ with $K$ lies in $K$. Hence, acting on the highest weight of $R$ with the generators of $iso(d-2)$, we get an irrep $R'$ of $iso(d-2)$ with $K$ represented by $0$. By construction, $R'$ is also an irrep of $so(1,1)\times so(d-2)$.

With that in mind, let us decompose $R$ under $so(1,1)\times so(d-2)$
\begin{equation}
R = \bigoplus_i (\lambda_i)\otimes R_i
\label{Rdecomp}\end{equation}
where $R_i$ is an irrep of $so(d-2)$ and $\lambda_i$ is the $so(1,1)$ weight labeling the corresponding 1-dimensional irrep of $so(1,1)$. Without loss of generality, we can order
\begin{displaymath}
\lambda_1 \gt \lambda_2 \geq \lambda_3 \geq \dots
\end{displaymath}
The embedding $so(1,1)\times so(d-2)\hookrightarrow so(1,d-1)$ is the one obtained by omitting the left-most node of the Dynkin diagram.
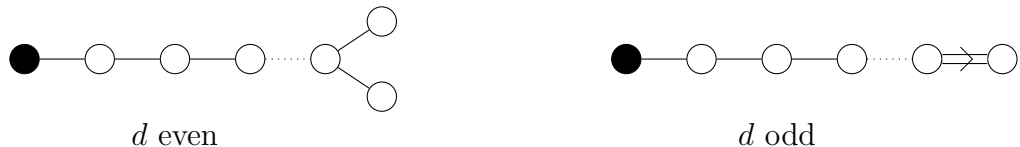
\begin{center}
\begin{tikzpicture}
\tikzset{->-/.style={decoration={
  markings,
  mark=at position .75 with {\arrow[scale=0.5]{Straight Barb}}},postaction={decorate}}}
\node[draw,circle,fill=black] (A) at (0,1) {};
\node[draw,circle,fill=white] (B) at (1,1) {};
\node[draw,circle,fill=white] (C) at (2,1) {};
\node[draw,circle,fill=white] (D) at (3,1) {};
\node[draw,circle,fill=white] (E) at (4,1) {};
\node[draw,circle,fill=white] (F) at (4.75,.5) {};
\node[draw,circle,fill=white] (G) at (4.75,1.5) {};
\node[draw,circle,fill=black] (AA) at (8,1) {};
\node[draw,circle,fill=white] (AB) at (9,1) {};
\node[draw,circle,fill=white] (AC) at (10,1) {};
\node[draw,circle,fill=white] (AD) at (11,1) {};
\node[draw,circle,fill=white] (AE) at (12,1) {};
\node[draw,circle,fill=white] (AF) at (13,1) {};
\node at (2,0) {$d$ even};
\node at (10,0) {$d$ odd};
\draw (A) -- (B);
\draw (B) -- (C);
\draw (C) -- (D);
\draw[dotted] (D) -- (E);
\draw (E) -- (F);
\draw (E) -- (G);
\draw (AA) -- (AB);
\draw (AB) -- (AC);
\draw (AC) -- (AD);
\draw[dotted] (AD) -- (AE);
\draw[double distance=3pt,->-] (AE) -- (AF);
\end{tikzpicture}
\end{center}
The remaining nodes are the simple roots of $so(d-2)$. The highest weight of $R$ under $so(1,d-1)$ is also the highest weight under the $so(1,1)\times so(d-2)$ subalgebra. That is, $R'$ is the representation $(\lambda_1)\otimes R_1$ in \eqref{Rdecomp}.

To be more explicit, we need some notation. Highest weight representations, with highest weight $\Lambda$, will be denoted by their Dynkin labels,
\begin{displaymath}
n_i = \frac{2 (\Lambda,\alpha_i)}{(\alpha_i,\alpha_i)}
\end{displaymath}
where the $\alpha_i$ are the simple roots.
Our convention for the $so(1,1)$ weights will be that the adjoint representation of $so(1,d-1)$ decomposes as
\begin{equation}\label{adjdecomp}
\begin{split}
[0,1,0,0,\dots,0] = (2)\otimes [1,0,0,\dots,0]\; &\oplus\; (0)\otimes [0,1,0,\dots,0] \\&\oplus\; (0)\otimes [0,0,0,\dots,0]\; \oplus\; (-2)\otimes [1,0,0,\dots,0]
\end{split}
\end{equation}
Here
\begin{itemize}%
\item $K=(2)\otimes [1,0,0,\dots,0]$. \emph{I.e.}, the generators of $K$ transform as a vector of $so(d-2)$ and with weight $+2$ under $so(1,1)$.
\item $(0)\otimes [0,1,0,\dots,0]$ is the adjoint of $so(d-2)$ and
\item $(0)\otimes [0,0,0,\dots,0]$ is the generator of $so(1,1)$.
\end{itemize}
The normalization of the $so(1,1)$ weights $(\lambda)$ is such that tensorial representations have $\lambda$ even and spinorial representations have $\lambda$ odd.

Let
\begin{equation}
R = [n,n_1,n_2,\dots,n_r]
\label{Rdef}\end{equation}
be our chosen highest weight representation of $so(1,d-1)$. The simple roots of $so(d-2)$ were obtained by omitting the first simple root. The corresponding Dynkin labels are obtained by omitting the first Dynkin label of $R$. The highest-weight of the $so(1,d-1)$ irrep $R$, with Dynkin labels \eqref{Rdef}, is the highest weight of the $so(d-2)$ irrep with Dynkin labels $[n_1,n_2,\dots,n_r]$. That is, our sought-after representation of $so(d-2)$ is
\begin{equation}
R_1 = [n_1,n_2,\dots,n_r]
\label{R1def}\end{equation}
Though we don't need it, the $so(1,1)$ weight is also determined:
\begin{equation}
\lambda_1 = \begin{cases}
2n + n_r +        2{\displaystyle \sum_{i=1}^{r-1}} n_i&\quad d=2r+3\\
2n + n_r + n_{r-1} + 2{\displaystyle \sum_{i=1}^{r-2}} n_i&\quad d=2r+2
\end{cases}
\label{lambda1def}\end{equation}
Finally, let us translate Weinberg's Young diagrams into the corresponding Dynkin labels of irreps. Consider a Young diagram, whose rows have lengths $l_0,l_1,\dots l_r$, where $r=(d-2)/2$ for $d$ even and $(d-3)/2$ for $d$ odd.

For $d$ odd, the corresponding Dynkin labels for $R$ are
\begin{align*}
n &= l_0- l_1\\
n_i &= l_i-l_{i+1}, \quad i=1,\dots,r-1\\
n_r &= 2 l_r\\
\intertext{For $d$ even,}
n &= l_0- l_1\\
n_i &= l_i-l_{i+1}, \quad i=1,\dots,r-2\\
n_{r-1}+ n_r &= 2 l_{r-1}\\
|n_{r-1} - n_r| &= 2 l_r
\end{align*}
Note that (of course) we only get tensorial representations this way ($n_r =\text{even}$ for $d$ odd or $n_{r-1}+n_r = \text{even}$ for $d$ even). Moreover, when $d$ is even and $l_r \gt 0$, the Young diagram corresponds to a \emph{reducible} representation, decomposing into \emph{two} irreps whose Dynkin labels differ by exchanging $n_{r-1}\leftrightarrow n_r$.

For tensorial representations, dropping the first Dynkin label in passing from $R$ to $R_1$ is precisely the same as Weinberg's conjectured ``decapitation'' procedure: removing the first row of the Young diagram. But it extends naturally to spinorial representations as well. And, for $d$ even, it takes care of the reducibility of Young diagrams with $l_r \gt 0$. Finally, it gives an interpretation of the $so(1,1)$ weight in \eqref{lambda1def}: $\lambda_1=2l_0$, where $l_0$ is the length of the row that he removes.

\section*{{Acknowledgements}}\label{acknowledgements}

Thanks to Steven Weinberg for posing the question answered herein and for insisting on clarity of the response. Any residual lack of clarity is the fault of the author. Thanks, also, to Yale Fan for helpful comments. This work was supported by the National Science Foundation under grant number PHY--1914679.

\bibliographystyle{utphys}

\bibliography{ref}

\end{document}